\documentclass[letterpaper]{article} 
\usepackage{aaai2027}  
\usepackage[hyphens]{url}  
\usepackage{graphicx} 
\usepackage{natbib}  
\usepackage{caption} 
\usepackage{amsmath}
\usepackage{multirow}
\usepackage{makecell}

\usepackage{booktabs}

\title{AgentCompile: An LLM-Guided Compiler for Direct CUDA Inference}
\author {
    Xuanzhe Li\textsuperscript{\rm 1,\rm 2}\equalcontrib,
    Ziyan Weng\textsuperscript{\rm 1}\equalcontrib,
    Zhiyu Zhu\textsuperscript{\rm 1,\rm 2}\equalcontrib,
    Junhui Hou\textsuperscript{\rm 2}\corresponding
}
\affiliations {
    \textsuperscript{\rm 1}City University of Hong Kong (Dongguan)\\
    \textsuperscript{\rm 2}City University of Hong Kong\\
    lixuanzhede29@126.com, 72510283@cityu-dg.edu.cn, zhiyu.zhu@cityu-edu.dg.cn, jh.hou@cityu.edu.hk
}

\begin{document}

\maketitle

\begin{abstract}
Transformer inference increasingly relies on specialized compiler and runtime support, while recent LLMs can generate nontrivial CUDA kernels. However, unconstrained generation guarantees neither correctness nor performance. We present \textsc{AgentCompile}, an LLM-guided CUDA inference compiler that combines two complementary uses of LLMs. First, the LLM provides advisory metadata for compiler-derived region summaries and bounded candidate spaces. The compiler then instantiates template-based CUDA candidates, validates correctness, selects implementations by measured latency, and falls back when specialization is unsupported or unprofitable. Second, under compiler-defined contracts, the LLM directly generates five classes of decode-critical kernels to accelerate inference, prompted by distilled optimization principles. \textsc{AgentCompile} integrates these kernels into a serving runtime with paged KV cache, continuous batching, preemption, chunked prefill, and bucketed full-step CUDA Graph replay. Across six evaluated model families, \textsc{AgentCompile} achieves speedups of \textbf{2.23--6.98$\times$} over PyTorch eager for single-request generation, and \textbf{1.04--1.16$\times$} over vLLM for both single-request generation and multi-request serving. We will open-source the project. 
\end{abstract}

\section{Introduction}

Modern transformer inference relies on compilers and runtimes such as PyTorch 2, FlashAttention, and vLLM~\citep{ansel2024pytorch,dao2022flashattention,kwon2023efficient}, which map model programs to efficient accelerator execution. Yet each path leaves a gap during autoregressive decoding: PyTorch eager retains substantial Python dispatch and dynamic-cache overhead; \texttt{torch.compile} suffers graph breaks, dynamic cache updates, and shape variation under long-context generation; serving runtimes such as vLLM rely on a fixed library of expert-written kernels. As our experiments show, these overheads compound to an end-to-end latency gap of up to ${\sim}7\times$ for single-request generation. Directly asking an LLM to write the missing CUDA kernels is risky: correctness hinges on low-level details from indexing and synchronization to launch configuration, and even a correct kernel is useless if it is slower than the library kernel it replaces.

\begin{figure}[t]
\centering
\includegraphics[width=0.925\linewidth]{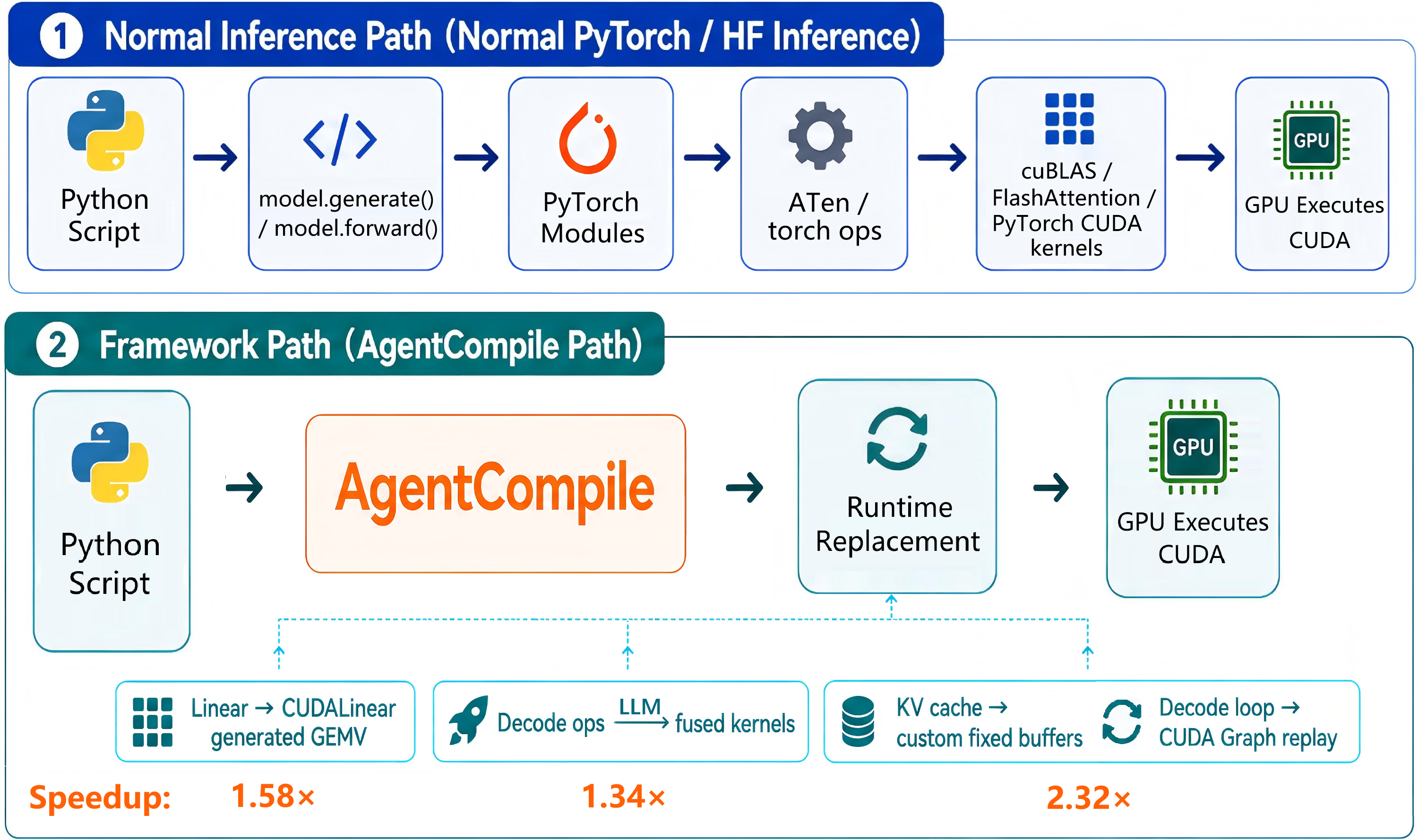}
\caption{Comparison of the standard PyTorch inference stack and \textsc{AgentCompile} for single-request inference.}
\label{fig:agentcompile_position}
\end{figure}

We find, however, that these risks can be contained. When generation is driven by distilled optimization principles rather than reference source code, and every candidate must pass compiler-side gates, an explicit performance-acceptance threshold, and in-graph measured selection before deployment, LLM-authored kernels can match or exceed expert-written and library kernels on real decode workloads, while a fallback path guarantees that the system never regresses when they do not.

We present \textsc{AgentCompile}, a compiler framework that combines two uses of LLMs. In the compiler pipeline, the LLM only advises: it labels compiler-constructed regions and prioritizes bounded template candidates, while the compiler generates and checks all code. On top of the pipeline, the LLM directly authors complete CUDA kernels for five decode-critical fusion patterns that span multiple compiler regions, which is beyond what template-based generation can express, and the compiler's role shrinks to checking: validation, measurement, and admission. This replaces what were previously hand-written, model-specific fused kernels and attention bodies, which rarely transferred across architectures, with kernels re-derived per model from the same principle libraries, so that supporting a new model requires generation rather than re-engineering. Both tiers feed a serving-grade runtime with paged KV cache, continuous batching, and full-step CUDA Graph replay. Figure~\ref{fig:agentcompile_position} positions \textsc{AgentCompile} as a substitution layer between Python model code and CUDA execution.

This paper makes three contributions: (i) a compiler-controlled pipeline that maps LLM advice to template-generated CUDA candidates; (ii) a principle-driven LLM kernel-generation methodology; and (iii) a serving-grade hybrid runtime evaluated across six model families including a vision--language model, outperforming PyTorch eager and \texttt{torch.compile} on single-request generation, and vLLM on both single-request generation and multi-request serving.

\section{Related Work}

\textbf{Neural-network compilers:} Compiler stacks and auto-schedulers lower models to accelerators through templates, schedules, and rewrites~\citep{chen2018tvm,openxla2024xla,ansel2024pytorch,lattner2021mlir,tillet2019triton,chen2018learning,zheng2020ansor,zhu2022roller,shi2023welder,jia2019taso,niu2021dnnfusion,zheng2022astitch,ma2020rammer}; \textsc{AgentCompile} is complementary, adding an LLM-authored kernel tier on top of compiler-defined constraints, validation, and measurement. \textbf{Inference runtimes and kernels:} Serving runtimes and specialized kernels~\citep{kwon2023efficient,yu2022orca,nvidia2024cudagraphs,dao2022flashattention} optimize scheduling and decode execution; \textsc{AgentCompile} implements the same serving-grade mechanisms but sources its decode kernels from contract-gated LLM generation rather than a fixed expert-written library. \textbf{LLMs for code and compiler optimization:} Unlike direct LLM kernel generation as evaluated by KernelBench~\citep{ouyang2025kernelbench} and related LLM-for-code work~\citep{chen2021evaluating,li2022competition,li2023starcoder,shinn2023reflexion,ni2024next,cummins2023large}, \textsc{AgentCompile} renders its prompts from distilled principle libraries (including falsified dead ends) rather than task descriptions or reference source, and every kernel must pass compiler contracts, acceptance thresholds, and runtime measured admission with a never-worse fallback. An extended discussion of related work is provided in the technical appendix.

\section{Problem Formulation and Design Goals}
\label{Sec:Problem}

Existing CUDA-based inference pipelines are built from fixed, hand-engineered kernels and hard-wired execution paths, so they offer little customization for a given model and cannot adapt their kernels to the structure of the network being served. Handing the task to an LLM does not by itself resolve this: freely generated CUDA code offers no guarantee of correctness or speed, and an unverified kernel cannot be trusted inside a production decode loop. The problem is therefore to admit LLM output, whether advisory metadata or complete kernel source, into an inference compiler such that it becomes executable only after passing compiler-controlled contracts, empirical validation, and measured selection.

We model an inference program as a directed graph $G=(V,E)$ of tensor operations and data dependencies. A region $r\subseteq V$ is a connected subgraph with an external interface (tensors, dependencies, dtype/layout assumptions, reduction axes, calling convention); a CUDA replacement is admissible only if it preserves this interface and passes the validation protocol of the surrounding inference path.

We decompose the problem into two tiers of LLM involvement, both governed by a shared admission discipline: Tier 1 uses the LLM only as an advisor over compiler-enumerated template candidates, Tier 2 lets the LLM author complete kernels under compiler-defined contracts, and a runtime admission layer decides by measurement whether any validated kernel actually runs.

\paragraph{Tier 1: advisory search over template candidates.}
For a region $r$ and hardware configuration $H$, the compiler constructs an executable candidate space $\mathcal{C}(r,H)$ of template implementations with realizable parameters (full list in the appendix). The LLM observes a compiler-generated region summary $S(r)$ rather than raw source code and returns advisory metadata
\[
\mathcal{A}_{\mathrm{LLM}}(S(r),\mathcal{C}(r,H))=(\ell_r,\pi_r,\eta_r,\rho_r),
\]
where $\ell_r$ is a semantic label, $\pi_r$ is a candidate preference order, $\eta_r$ contains parameter hints, and $\rho_r$ contains risk annotations, all of which are treated as hypotheses, with unsupported labels, invalid parameters, and boundary-crossing suggestions rejected by compiler checks.

\paragraph{Tier 2: contract-gated kernel generation.}
For a recognized fusion pattern $p$ (the five families detailed in Method), the compiler defines a contract $\Sigma(p)$, which specifies the tensor interface, dtype rules, numerical tolerances, and a performance floor, together with a principle library $P(p)$ of distilled optimization knowledge that contains no kernel source. The LLM produces complete kernel source $k=\mathrm{LLM}(P(p),\Sigma(p))$, which enters the validated set $\mathcal{V}(p)$ only if it compiles, satisfies the interface and dtype rules, matches reference outputs within tolerance, and reaches a measured acceptance threshold against the existing implementation; generation runs multiple rounds with structured failure feedback, keeping the best candidate.

\paragraph{Runtime admission and never-worse selection.}
At runtime, each validated kernel is timed \emph{in-graph} against the existing implementation $b$ and selected only if $T_{\mathrm{graph}}(k)\le\mu\,T_{\mathrm{graph}}(b)$, with a small margin $\mu$ (we use $\mu{=}1.04$) absorbing measurement noise; anything slower falls back to the hand-written or library path, and among validated candidates the runtime selects $c^\star=\arg\min_{c\in\mathcal{V}}T(c)$, so end-to-end performance is never worse than the baseline path by more than the margin. Validation is empirical rather than formal equivalence: it uses numerical comparison under dtype-appropriate tolerances when a reference exists, and structural and end-to-end consistency checks otherwise.

\section{Method}
\label{Sec:Method}

\begin{figure*}[t]
  \centering
  \includegraphics[width=0.82\linewidth]{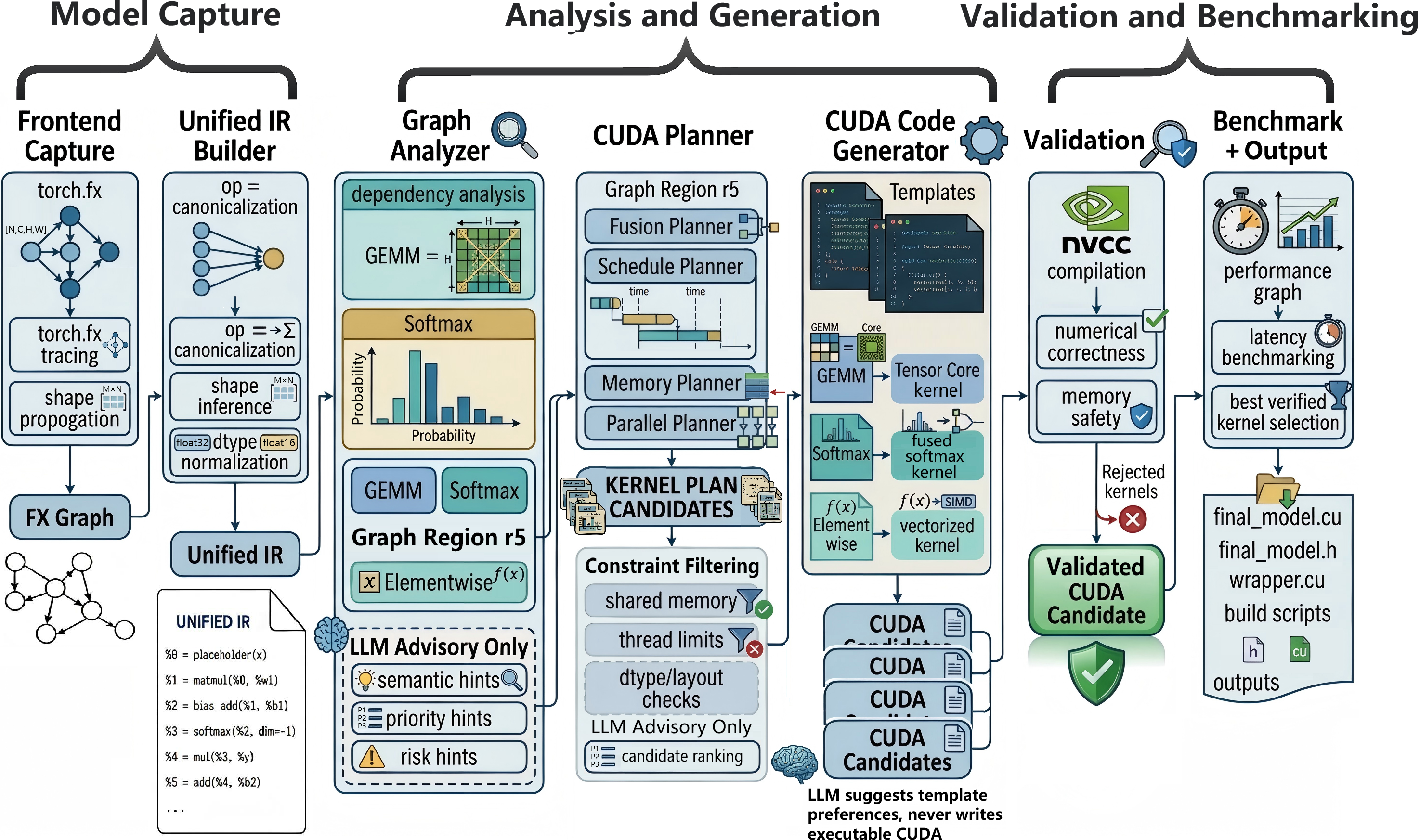}
    \caption{
  Running example of compiler-guided LLM assistance on a representative region r5. The Graph Analyzer derives a structured region summary for the LLM. The LLM may identify the region as a softmax or normalization pattern, and return semantic labels, candidate preferences, parameter and risk hints (considered hypotheses rather than correctness guarantees); the Planner checks dependencies, reduction semantics, dtype/layout constraints, template availability, and hardware limits before any CUDA is materialized; only plans that pass constraint filtering are sent to the Code Generator and instantiated as CUDA candidates. Once they pass validation checks, the best validated kernels are selected as the final CUDA kernel outputs.
  }
  \label{fig:running-example}
\end{figure*}

\begin{figure}[t]
  \centering
  \includegraphics[width=1\linewidth]{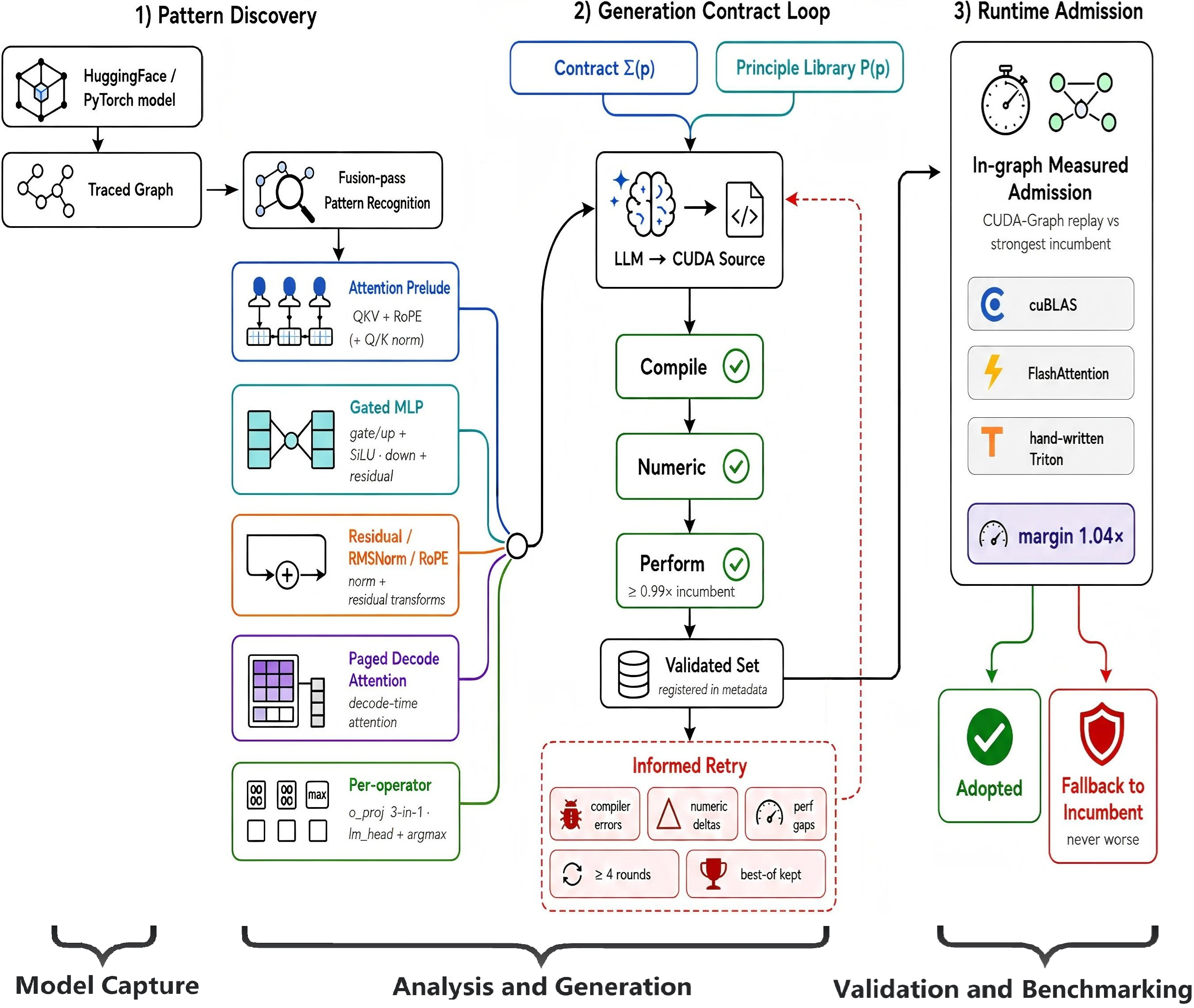}
  \caption{Generation pipeline of the five LLM-generated kernel families: recognized patterns branch into per-family contracts $\Sigma(p)$ and source-free principle libraries $P(p)$; the LLM produces complete CUDA source that must pass the compile, numeric, and performance gates, with informed retry on structured failure feedback; validated kernels reach production only through in-graph measured admission, and losers fall back. Failure modes flow back into $P(p)$ as falsification clauses; the decode-attention kernel propagates across models as a re-validated seed.}
  \label{fig:llm-author-kernels}
\end{figure}

\textsc{AgentCompile} implements the formulation above as a three-stage compiler pipeline, plus a serving-grade runtime layer. The pipeline captures a PyTorch or HuggingFace model, forms compiler regions, constructs bounded CUDA candidate spaces, and selects executable template kernels through validation and measurement, where the compiler generates and checks all code and the LLM only advises. The LLM kernel generator then directly authors the five decode-critical kernel families under compiler-defined contracts, where the compiler only checks to ensure correctness.
Figure~\ref{fig:running-example} illustrates the compiler pipeline's key contract on a representative region r5; Figure~\ref{fig:llm-author-kernels} illustrates how the five kernel families are generated by the LLM kernel generator.

\paragraph{Pipeline Overview.}
\textsc{AgentCompile} organizes the pipeline into three stages: \textbf{model capture}, \textbf{analysis \& generation}, and \textbf{validation \& benchmarking}.

\paragraph{Model capture:} reads the target model code and obtains a traced computation graph with example inputs, tensor shapes, dtypes, layouts, and dependency metadata~\citep{ansel2024pytorch,MLSYS2022_7c98f9c7}.
\textbf{Analysis \& generation:} converts the traced graph into CUDA candidates. The graph and its shape, dtype, layout, dependency, and aliasing metadata are given to the \textit{Graph Analyzer}
, which partitions the computation into candidate regions and records each region's operator sequence, tensor interfaces, reduction axes, dominant operation type, memory-access pattern, and dynamic-shape flags. The \textit{Planner} 
then enumerates legal fusion, schedule, memory, and parallelization choices for each region. These bounded kernel plans, together with their region interfaces and template families, are passed to the \textit{CUDA Code Generator} 
to instantiate CUDA candidates. 
\textbf{Validation \& benchmarking:} compiles, validates, benchmarks, and packages the generated kernels, following the common compile--measure--select pattern~\citep{chen2018tvm,zheng2020ansor}. Invalid candidates are rejected by compilation, interface, and numerical checks when a reference path is available; the fastest validated implementation is selected; and the system outputs CUDA kernel source files, wrappers, headers, metadata, and build scripts.

\noindent\textbf{Model capture} and \textbf{validation \& benchmarking} are standard, well-studied components that can be built with straightforward graph-capture, validation, benchmarking, and packaging logic. We therefore focus on the \textbf{Analysis \& generation} stage, 
which is the part that most distinguishes \textsc{AgentCompile}.

\subsection{Graph Analyzer: Region Construction}
\label{Sec:GraphAnalyzer}
The analyzer determines which operations may form a candidate region before LLM consultation, as illustrated in Figure~\ref{fig:running-example}. Traversing the IR (intermediate representation) along use-def dependencies, it merges consecutive pointwise, activation, softmax, norm, and reduction nodes under compatible dtype/layout assumptions, treats layout-only nodes as boundary metadata, keeps matmuls as independent regions (attaching simple epilogues when constraints allow), and makes in-place operations, side effects, unresolved aliasing, unsupported dtypes, and ambiguous dynamic shapes hard boundaries. Each accepted region carries its operator sequence, shapes, dtypes, reduction axes, and access pattern, and rule-based recognition assigns semantic categories; the same layer detects the five fusion patterns handed to the LLM kernel generator, using structural preconditions as hard gates so that models violating a precondition are safely skipped for that pattern. Within the analyzer, the LLM is invoked only where semantic analysis is useful (mixed regions, complex GEMM (general matrix--matrix multiplication) epilogues, reduction--pointwise regions), and region boundaries are accepted only after deterministic dependency, alias, and side-effect checks.

\subsection{CUDA Planner}
\label{Sec:Generator}
For the template-based generation path illustrated in Figure~\ref{fig:running-example}, the planner converts each region into a bounded set of kernel plans, rather than asking the LLM to invent implementations: a fusion choice (single-kernel, full, partial, none), a template family (tiled or Tensor-Core GEMM, softmax, reductions, normalization, elementwise, decode-oriented GEMV (general matrix--vector multiplication)), and memory/parallel parameters (vector width, shared-memory policy, grid/block shape, warp mapping); constraint filtering and deterministic ranking then bound the search. For high-value regions, the LLM may reorder top candidates or suggest parameters, which are accepted only if they already belong to the enumerated plan space, and a diversity filter preserves multiple template families.

\subsection{CUDA Code Generator}
\paragraph{Template-Based Code Generator.} The template-based generator in Figure~\ref{fig:running-example} materializes checked plans from deterministic templates, which specify all executable details (indexing, boundary checks, dtype conversion, synchronization, and launch signatures). Decode-oriented GEMV matters most for autoregressive decoding: weights are stored in an $[N,K]$ layout for contiguous row reads, blocks use 128-bit vectorized loads and warp-shuffle reductions, and a dual-path adaptive-$N$ strategy keeps enough blocks resident for large $N$.

\paragraph{Principle-Driven LLM Kernel Generator.} To complement kernel generation, this generator uses LLM to author complete CUDA kernels for the five recognized fusion patterns. Figure~\ref{fig:llm-author-kernels} details this kernel-authoring workflow, whose components are described below.

The fusion pass routes recognized patterns to five LLM-generated kernel families that together cover the decode-critical path of a transformer layer:
(i) \emph{attention-prelude fusion}: QKV projection + RoPE (rotary position embedding), optionally with per-head Q/K RMSNorm;
(ii) \emph{gated-MLP fusion}: gate/up projection + SiLU and down projection + residual add, each in GEMV and thin-GEMM variants dispatched by batch size;
(iii) \emph{residual--norm/RoPE fusion}: residual add + RMSNorm, plus a standalone RoPE kernel;
(iv) \emph{paged decode attention}: the attention body itself, an online-softmax kernel over paged KV with fused cross-CTA (cooperative thread array) merging;
(v) \emph{per-operator decode fusion}: an output-projection (\texttt{o\_proj}) kernel fusing GEMV + residual + post-RMSNorm, and a language-model-head (\texttt{lm\_head}) + argmax kernel producing token ids without materializing the logits.
Every generated kernel competes against the strongest existing implementation at its site (cuBLAS, FlashAttention, or hand-written Triton) under the measured-admission rule; encoder-only models reuse the same machinery through a parallel encoder-kernel family.

To constrain the generation process, we
introduce a principle library, in which the prompts are rendered from a per-pattern library of distilled optimization principles plus a performance mandate. The library contains \emph{no kernel source}: each block records design decisions and, crucially, \emph{falsified dead ends}, that is, approaches that were tried and measured to fail, so that the LLM does not rediscover them. Principles are dimension- and model-agnostic, so the same library serves every captured model and a kernel validated for one model can seed another.

When the principle library cannot cover the pattern at the required level, we hand-write a reference kernel, distill its design decisions and dead ends into prose, and let the LLM re-derive a kernel without ever seeing the reference source. For paged decode attention, the hand-written reference reaches 0.967--1.00$\times$ of FlashAttention at ${\sim}90\%$ of A800 peak memory bandwidth (details in the appendix); the LLM-generated kernel reaches 0.966$\times$, within 0.1\% of the reference.

Each pattern is generated with acceptance threshold $\theta_{\mathrm{gen}}{=}0.99$ against the existing implementation, at least $N{=}4$ rounds keeping the best candidate, and informed retry over structured diagnostics (compilation errors, numerical mismatches, slow measurements). Contract clauses are exhaustive about dtypes, since a single unstated dtype assumption can cause every round to fail, and unprofitable sub-families are early-stopped.

\subsection{Kernel Customization and Multi-Request Adaptation} 
\label{Sec:Runtime}

We further introduce several kernel customizations, including the custom linear layer, KV cache, and attention, to boost the efficiency of the proposed infrastructure. Meanwhile, to adapt the method to real serving applications, we design serving-level components ranging from globally asynchronous prefill--decode steps to a locally customized paged KV cache mechanism.

\begin{table*}[!t]
\centering
\small
\setlength{\tabcolsep}{3.5pt}
\begin{tabular}{llcccccc}
\toprule
\multirow{2}{*}{Model} &
\multirow{2}{*}{Framework} &
\multicolumn{5}{c}{Workload (input/output)} &
\multirow{2}{*}{Avg. Speedup} \\
\cmidrule(lr){3-7}
& & i128/o128 & i2048/o512 & i8192/o4096 & i12288/o8192 & i20480/o16384 & \\
\midrule
\multirow{3}{*}{Qwen3-4B}
& PyTorch & 4765 & 18974 & 151807 & 303239 & 603417 & -- \\
& \texttt{torch.compile} & 1707/2.8$\times$ & 7092/2.7$\times$ & 76237/2.0$\times$ & 202430/1.5$\times$ & 602062/1.0$\times$ & 1.99$\times$ \\
& Ours & 824/5.8$\times$ & 3482/5.4$\times$ & 30925/4.9$\times$ & 66095/4.6$\times$ & 150819/4.0$\times$ & \textbf{4.95$\times$} \\
\midrule
\multirow{3}{*}{Falcon3-7B}
& PyTorch & 2880 & 11694 & 91589 & 186789 & -- & -- \\
& \texttt{torch.compile} & 1457/2.0$\times$ & 6809/1.7$\times$ & 81197/1.1$\times$ & 201527/0.9$\times$ & -- & 1.44$\times$ \\
& Ours & 1194/2.4$\times$ & 5089/2.3$\times$ & 43022/2.1$\times$ & 90409/2.1$\times$ & -- & \textbf{2.23$\times$} \\
\midrule
\multirow{3}{*}{Mistral-7B}
& PyTorch & 3164 & 12822 & 100928 & 203672 & -- & -- \\
& \texttt{torch.compile} & 1510/2.1$\times$ & 7401/1.7$\times$ & 94048/1.1$\times$ & 240877/0.8$\times$ & -- & 1.44$\times$ \\
& Ours & 1212/2.6$\times$ & 5086/2.5$\times$ & 43130/2.3$\times$ & 89895/2.3$\times$ & -- & \textbf{2.43$\times$} \\
\midrule
\multirow{3}{*}{GLM-4-9B}
& PyTorch & 5023 & 19966 & 161703 & 320419 & 640323 & -- \\
& \texttt{torch.compile} & 2105/2.4$\times$ & 8619/2.3$\times$ & 95306/1.7$\times$ & 233744/1.4$\times$ & 641812/1.0$\times$ & 1.75$\times$ \\
& Ours & 1668/3.0$\times$ & 6856/2.9$\times$ & 57668/2.8$\times$ & 115555/2.8$\times$ & 239863/2.7$\times$ & \textbf{2.83$\times$} \\
\midrule
\multirow{3}{*}{Qwen3-VL-2B}
& PyTorch & 3575 & 14181 & 113733 & 229348 & 456021 & -- \\
& \texttt{torch.compile} & 1244/2.9$\times$ & 5093/2.8$\times$ & 41139/2.8$\times$ & 103825/2.2$\times$ & 303122/1.5$\times$ & 2.43$\times$ \\
& Ours & 416/8.6$\times$ & 1772/8.0$\times$ & 16626/6.8$\times$ & 36562/6.3$\times$ & 87523/5.2$\times$ & \textbf{6.98$\times$} \\
\bottomrule
\end{tabular}
\caption{Representative single-request end-to-end generation latency (ms) and speedup ratio measured against PyTorch eager execution. Workload denotes input/output length; `--' marks workloads exceeding the model's maximum context length. The last column reports the average speedup over the listed workloads.}
\label{tab:e2e-representative-main}
\end{table*}

\begin{table*}[!t]
\centering
\small
\setlength{\tabcolsep}{3.5pt}
\begin{tabular}{llcccccc}
\toprule
\multirow{2}{*}{Model} &
\multirow{2}{*}{Framework} &
\multicolumn{5}{c}{Workload (input/output)} &
\multirow{2}{*}{Avg. Speedup} \\
\cmidrule(lr){3-7}
& & i128/o128 & i2048/o512 & i8192/o4096 & i12288/o8192 & i20480/o16384 & \\
\midrule
\multirow{2}{*}{Qwen3-4B}
& vLLM & 905 & 3727 & 32690 & 69769 & 158059 & -- \\
& Ours
& 824/1.10$\times$ & 3482/1.07$\times$ & 30925/1.06$\times$ & 66095/1.06$\times$ & 150819/1.05$\times$ & \textbf{1.07$\times$} \\
\midrule
\multirow{2}{*}{Llama-3.2-3B}
& vLLM & 726 & 2994 & 26125 & 55615 & 125507 & -- \\
& Ours
& 652/1.11$\times$ & 2749/1.09$\times$ & 24321/1.07$\times$ & 51968/1.07$\times$ & 118497/1.06$\times$ & \textbf{1.08$\times$} \\
\midrule
\multirow{2}{*}{Falcon3-7B}
& vLLM & 1329 & 5496 & 46082 & 96415 & -- & -- \\
& Ours
& 1194/1.11$\times$ & 5089/1.08$\times$ & 43022/1.07$\times$ & 90409/1.07$\times$ & -- & \textbf{1.08$\times$} \\
\midrule
\multirow{2}{*}{Mistral-7B}
& vLLM & 1346 & 5557 & 46788 & 97519 & -- & -- \\
& Ours
& 1212/1.11$\times$ & 5086/1.09$\times$ & 43130/1.08$\times$ & 89895/1.08$\times$ & -- & \textbf{1.09$\times$} \\
\midrule
\multirow{2}{*}{Qwen3-VL-2B}
& vLLM & 508 & 2103 & 18997 & 41264 & 96700 & -- \\
& Ours
& 416/1.22$\times$ & 1772/1.19$\times$ & 16626/1.14$\times$ & 36562/1.13$\times$ & 87523/1.10$\times$ & \textbf{1.16$\times$} \\
\bottomrule
\end{tabular}
\caption{Representative single-request end-to-end generation latency (ms) and speedup ratio measured against vLLM.} 
\label{tab:e2e-representative-vllm}
\end{table*}

\begin{figure*}[!t] 
  \centering
  \includegraphics[width=0.9\linewidth]{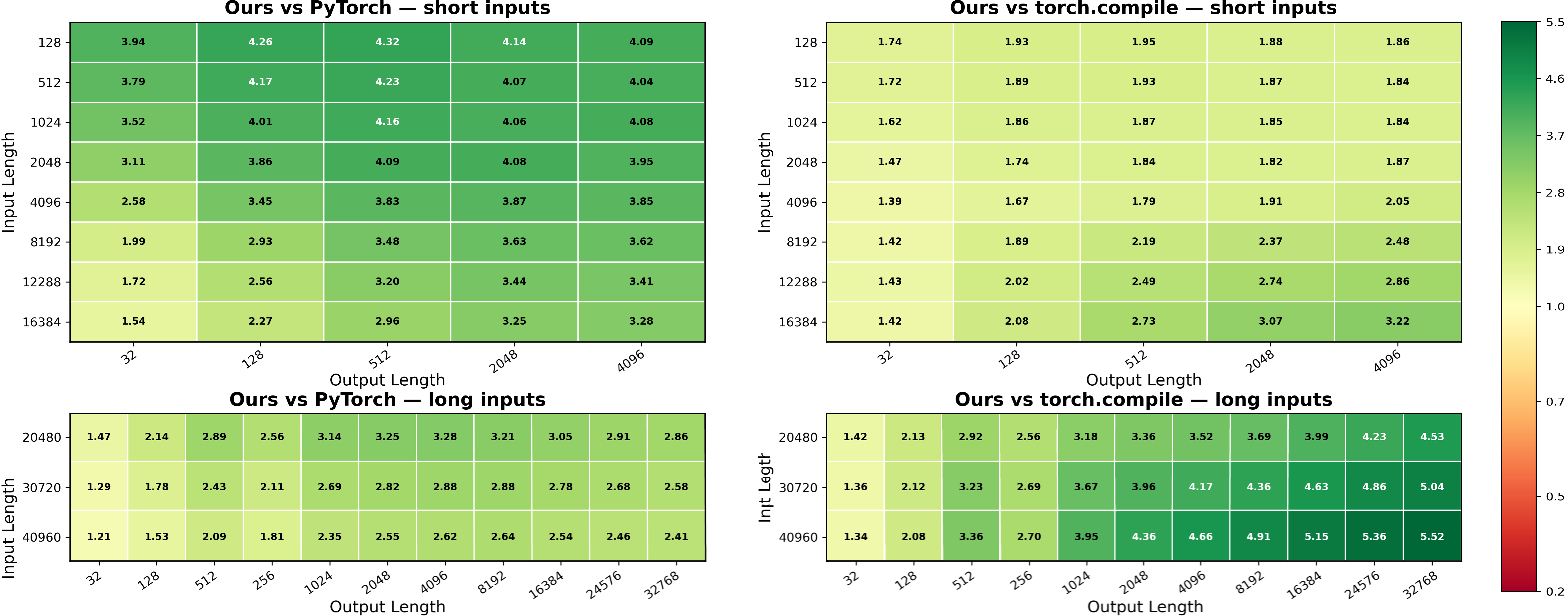}
  \caption{Input/output-length scaling comparison on Llama-3.2-3B-Instruct across PyTorch eager, \texttt{torch.compile}+FlashAttention, and \textsc{AgentCompile}.}
  \label{fig:speed-comparison-main}
\end{figure*}

\texttt{CUDALinear} replaces PyTorch \texttt{nn.Linear} on the decode path, dispatching to custom GEMM/GEMV kernels through a C++ extension with pre-transposed weights and an $M{=}1$ GEMV path for single-token decoding; prefill and unsupported shapes fall back to cuBLAS.
\texttt{FlashKVCacheManager} preallocates fixed-size K/V buffers, tracks per-layer sequence lengths on the GPU, and imports prefill states from the framework cache; attention runs through FlashAttention's KV-cache decode API or the generated paged decode-attention kernel. One exception is the KV append: FlashAttention's in-kernel bfloat16 (bf16) append is ${\sim}4\times$ slower than fp16 (0.73 vs.\ 0.18\,ms per decode step), so a small scatter \texttt{kv\_append} kernel writes the new K/V vectors and FlashAttention takes its no-append fast path with the sequence length advanced by one, which is semantically identical to the fused call, and bf16 generation becomes slightly faster than fp16 as a result.

\begin{table*}[!t]
\centering
\small
\setlength{\tabcolsep}{3.5pt}
\begin{tabular}{lcccccc}
\toprule
\multirow{2}{*}{Model} &
\multicolumn{5}{c}{Workload (input/output)} &
\multirow{2}{*}{Avg. Speedup} \\
\cmidrule(lr){2-6}
& i2k/o1k n32m32 & i2k/o4k n64m64 & i4k/o1k n32m32 & i8k/o8k n40m40 & i10k/o10k n30m30 & \\
\midrule
Qwen3-4B & 1.02$\times$ & 1.03$\times$ & 1.04$\times$ & 1.06$\times$ & 1.06$\times$ & \textbf{1.04$\times$} \\
Llama-3.2-3B & 1.03$\times$ & 1.04$\times$ & 1.05$\times$ & 1.04$\times$ & 1.27$\times$ & \textbf{1.09$\times$} \\
Mistral-7B & 1.05$\times$ & -- & 1.05$\times$ & 1.02$\times$ & -- & \textbf{1.04$\times$} \\
Qwen3-VL-2B & 1.04$\times$ & 1.06$\times$ & 1.06$\times$ & 1.09$\times$ & 1.34$\times$ & \textbf{1.12$\times$} \\
\bottomrule
\end{tabular}
\caption{Multi-request end-to-end throughput speedup over vLLM on the uniform suite. All requests in a tier share the same input/output lengths; tiers are labeled i$\langle$in$\rangle$/o$\langle$out$\rangle$ n$\langle n\rangle$m$\langle m\rangle$, where k${}={}$1024 tokens, $n$ is the total number of requests, and $m$ is the maximum number of concurrently admitted requests. `--' marks tiers exceeding the model's KV-memory budget.}
\label{tab:multibatch-uniform}
\end{table*}

\begin{table*}[!t]
\centering
\small
\setlength{\tabcolsep}{3.5pt}
\begin{tabular}{lccccccc}
\toprule
\multirow{2}{*}{Model} &
\multicolumn{6}{c}{Workload (KV pool $\times$ pressure)} &
\multirow{2}{*}{Avg. Speedup} \\
\cmidrule(lr){2-7}
& 12M & 16M & 60H & 30HH & 40HH & 60HH\\
\midrule
Qwen3-4B & 1.02$\times$ & 1.01$\times$ & 1.06$\times$ & 1.08$\times$ & 1.12$\times$ & 1.12$\times$ & \textbf{1.07$\times$} \\
Llama-3.2-3B & 1.05$\times$ & 1.02$\times$ & 1.03$\times$ & 1.13$\times$ & 1.12$\times$ & 1.09$\times$ & \textbf{1.07$\times$} \\
Falcon3-7B & 1.03$\times$ & 1.03$\times$ & 1.02$\times$ & 1.08$\times$ & 1.01$\times$ & 1.04$\times$ & \textbf{1.04$\times$} \\
Mistral-7B & 1.02$\times$ & 1.04$\times$ & 1.01$\times$ & 1.09$\times$ & 1.13$\times$ & 1.05$\times$ & \textbf{1.06$\times$} \\
Qwen3-VL-2B & 1.09$\times$ & 1.03$\times$ & 1.01$\times$ & 1.18$\times$ & 1.16$\times$ & 1.03$\times$ & \textbf{1.08$\times$} \\
\bottomrule
\end{tabular}
\caption{Multi-request end-to-end throughput speedup over vLLM on the ragged swap suite. Each tier issues a mixed-length request pool against a paged-KV pool, whose size is given by the number in the tier label (in GB, e.g., 12M runs against a 12GB pool). M (medium-pressure) tiers use requests that declare large output budgets but terminate early, so the declared load oversubscribes the KV pool by ${\sim}3\times$ and forces admission and preemption decisions; H (long) tiers use genuinely long mixed requests ($3\times$ length scale, prompts up to ${\sim}12$k tokens); HH (extreme) tiers scale request lengths $5$--$6\times$ (prompts up to ${\sim}24$k tokens). Both systems serve the identical request pool. Details in appendix.} 
\label{tab:multibatch-swap}
\end{table*}

\begin{table*}[!t] 
\centering
\small
\setlength{\tabcolsep}{3.5pt}
\begin{tabular}{lcccccc}
\toprule
\multirow{2}{*}{Framework Change}
& \multicolumn{5}{c}{Workload (i/o)}
& \multirow{2}{*}{Avg. Speedup} \\
\cmidrule(lr){2-6}
& i128/o128 & i2048/o512 & i8192/o4096 & i12288/o8192 & i20480/o16384 & \\
\midrule
baseline & 4765 & 18974 & 151807 & 303239 & 603417 & -- \\
+ CL \& GEMV \& StaticCache & 2986 & 11908 & 95645 & 192952 & 386487 & 1.58$\times$ \\
+ FAKV \& QKV\_proj \& Graph & 1160 & 4826 & 41245 & 86754 & 191991 & 2.32$\times$ \\
+ LLM \& LLM-generated kernels & 824 & 3482 & 30925 & 66095 & 150819 & 1.34$\times$ \\
= full system & -- & -- & -- & -- & -- & 4.95$\times$ \\
\bottomrule
\end{tabular}
\caption{Qwen3-4B stepwise runtime ablation for single-request autoregressive generation (batch size 1). Columns denote input/output sequence lengths, and values report mean latency in milliseconds. ``baseline'' denotes PyTorch eager with FlashAttention; ``CL'' denotes \texttt{CUDALinear}, ``GEMV'' denotes template-generated single-op GEMV kernels (compiler tier, not LLM-authored); ``FAKV'' denotes FlashAttention with KV-cache update and our KV-cache manager (including a custom bf16 \texttt{kv\_append} kernel replacing FlashAttention's slow bf16 append), ``QKV\_proj'' denotes fused QKV projection, ``Graph'' denotes CUDA Graph replay; ``LLM'' denotes the LLM acting as a semantic advisor to the compiler pipeline (labels and candidate hints, no kernel authoring); ``LLM-generated kernels'' denotes the five families of LLM-authored kernels.}
\label{tab:runtime-ablation}
\end{table*}

For the multi-request adaptation, we implement serving-grade mechanisms end to end. For memory, a paged KV cache allocates at block granularity, and requests are preempted by recomputation or by swapping KV blocks to host memory under an aggressive admission policy. For scheduling, the engine batches requests continuously and mixes chunked prefill with decode steps. For execution, decode steps replay as \emph{bucketed full-step CUDA Graphs}, where one graph per batch-size bucket covers the entire batched step, eliminating per-token launch and dispatch overhead; memory budgeting is parameterized per model.

\section{Experiments}

We evaluate \textsc{AgentCompile} on single-request end-to-end generation and multi-request serving, quantifying the acceleration over PyTorch eager, \texttt{torch.compile}, and vLLM; kernel-level quality, compiler-pipeline statistics, and encoder-only vision results are reported in the appendix. All LLM stages use Claude-Opus-4.8.

\noindent\textbf{Setup.}
Experiments use an NVIDIA A800 SXM4 80GB GPU with an Intel Xeon Platinum 8358 host. We evaluate six model families: Qwen3-4B, Llama-3.2-3B-Instruct, Falcon3-7B-Instruct, Mistral-7B-Instruct-v0.3, GLM-4-9B-chat, and Qwen3-VL-2B-Instruct (text stack), 
in bfloat16. 
The software stack, latency accounting, timing protocol, and the execution path of each system are detailed in the appendix. 

\subsection{Single-Request End-to-End Latency Comparisons}
\label{sec:e2e-results}

Figure~\ref{fig:speed-comparison-main} and Tables~\ref{tab:e2e-representative-main}--\ref{tab:e2e-representative-vllm} compare PyTorch eager, \texttt{torch.compile}+FlashAttention, vLLM, and \textsc{AgentCompile} across representative workloads: \textsc{AgentCompile} is 2.23--6.98$\times$ faster than PyTorch eager on average, consistently above \texttt{torch.compile} (1.44--2.43$\times$), whose advantage collapses on long workloads, and faster than vLLM on \emph{every} measured workload (1.07--1.16$\times$ on average).

Figure~\ref{fig:speed-comparison-main} shows that \textsc{AgentCompile}'s speedup over PyTorch eager generally becomes more pronounced as output length increases, especially on long-input workloads. This reflects the decode-oriented design of the runtime: paged KV management, full-step CUDA Graph replay, and generated GEMV/fused kernels reduce per-token overhead across long autoregressive loops. 

In contrast, PyTorch eager retains substantial framework dispatch and dynamic-cache overhead. 
\texttt{torch.compile} is effective on short-input workloads, where fusion and compiled graph fragments reduce framework and kernel-launch overhead; as input length increases, computation and KV cache traffic dominate, and graph breaks, dynamic cache updates, and shape changes further reduce graph reuse. At very long outputs, the speedup gradually decreases because the growing KV cache makes attention increasingly dominant and less differentiated across systems. Nevertheless, our \textsc{AgentCompile} still substantially outperforms both PyTorch eager and \texttt{torch.compile} by a wide margin in this regime.

The vLLM comparison in Table~\ref{tab:e2e-representative-vllm} should be interpreted carefully, because vLLM and \textsc{AgentCompile} optimize different and independent parts of the inference stack. 
vLLM is a serving-oriented runtime whose strengths lie in paged KV management, request scheduling, and production decode execution, whereas \textsc{AgentCompile}'s contribution is the kernel side: compiler-selected and LLM-generated kernels, fused decode components, and full-step CUDA Graph replay. The two systems thus improve largely complementary parts of the inference stack. 
In principle, the kernels studied here could also serve inside a scheduler such as vLLM's.

\subsection{Multi-Request Throughput vs.\ vLLM}
\label{sec:multibatch}

We further compare the continuous-batching engine against vLLM under a \emph{uniform} suite and a \emph{ragged swap} suite (Tables~\ref{tab:multibatch-uniform} and~\ref{tab:multibatch-swap}; workload construction in the captions). The engine is faster than vLLM on every runnable tier of both suites; the advantage widens on the long-output uniform tiers, where bucketed full-step graph replay amortizes per-step host overhead over many decode steps, and on the extreme long-context (HH) swap tiers, where aggressive admission with KV-swap preemption keeps the GPU saturated.

\subsection{Ablation Studies} 

Table~\ref{tab:runtime-ablation} reports a stepwise Qwen3-4B ablation. \texttt{CUDALinear} with single-op GEMV kernels and a static KV cache (replacing the PyTorch DynamicCache) yields 1.58$\times$ by removing framework and dynamic-cache overhead. Adding FAKV, fused QKV projection, and CUDA Graph replay is the largest step (2.32$\times$): FAKV is near-neutral in eager mode but provides the graph-capturable decode path that full-step replay requires; capturing a CUDA Graph on the SDPA (scaled dot-product attention) + StaticCache path instead plateaus at 1318\,ms on i128/o128, versus 824\,ms with FAKV (both measured in the full-system configuration). The LLM tier contributes a further 1.34$\times$, for 4.95$\times$ overall. Stage averages are per-workload means, so their product is not exactly equal to the full-system average, which is computed directly from the baseline and full-system latencies.

\section{Conclusion}

This paper presented \textsc{AgentCompile}, a compiler framework in which an LLM directly authors decode-critical CUDA kernels under compiler-defined contracts. The key idea is to separate authorship from admission: the LLM writes complete kernels from distilled optimization principles, while compilation, interface, numerical gates, 
and runtime in-graph measured selection with fallback decide what actually runs. Combined with a serving-grade runtime
, the system outperforms PyTorch eager and \texttt{torch.compile} on single-request generation, and vLLM on both single-request generation and multi-request serving. 
The methodology of hand-writing once, distilling to principles, and generating everywhere offers a practical path for accumulating kernel-engineering knowledge in a form that LLMs can reuse.

\section{Limitations}

\textsc{AgentCompile} is a prototype. It targets transformer-style inference regions, and unsupported regions fall back to the original framework or library path. To further improve the performance of the proposed method, we will explore more refined and efficient CUDA kernels in future work.

\appendix

\section{Supplementary Implementation Notes}
\label{app:implementation}

This appendix records implementation details that support, but do not repeat, the main method. The frontend supports single-file modules, project directories, and HuggingFace identifiers, and resolves model classes through the framework's authoritative configuration registry; command-line contracts require the model path, class, and identifier to be mutually consistent. The IR builder normalizes operator names and records numerical contracts such as precise softmax/norm behavior, associative reduction or matmul behavior, and relaxed elementwise behavior. The analyzer calls the LLM only for regions whose rule-based pattern is mixed, complex GEMM-epilogue, or reduction-pointwise and whose region size justifies semantic analysis. Accepted LLM metadata is stored as region hints and planner preferences; rejected advice is ignored without changing the graph. Generated fusion kernels are registered in emitted metadata under per-model name prefixes; standalone generator entry points can regenerate a single kernel class and merge its registration without re-running the full pipeline. The hand-written paged decode-attention reference distilled in the main text is a warp-autonomous Tensor-Core pipeline using \texttt{m16n8k16} \texttt{mma} fragments whose accumulator layout is lane-isomorphic to the operand layout, enabling in-fragment softmax, with \texttt{ldmatrix} loads, padded shared-memory rows to avoid bank conflicts, and a fused fence-based merge that removes a separate reduction kernel.

\begin{table*}[t]
\centering
\small
\caption{Compiler-pipeline statistics for the seven model families. Validated denotes candidates that passed compilation and numerical checks under the logged validation protocol; one implementation is selected per region.}
\label{tab:pipeline-evidence}
\begin{tabular}{lcccccc}
\toprule
Run & IR nodes & Regions & Plans / candidates & Validated & Selected & Emitted artifacts \\
\midrule
Qwen3-4B & 236 & 34 & 174 & 174/174 & 34 & 7 files / 43 signatures \\
Llama-3.2-3B-Instruct & 192 & 34 & 174 & 174/174 & 34 & 7 files / 43 signatures \\
Falcon3-7B-Instruct & 192 & 34 & 227 & 227/227 & 34 & 7 files / 43 signatures \\
Mistral-7B-Instruct-v0.3 & 201 & 37 & 249 & 249/249 & 37 & 7 files / 46 signatures \\
GLM-4-9B-chat & 238 & 50 & 347 & 347/347 & 50 & 7 files / 57 signatures \\
Qwen3-VL-2B-Instruct & 291 & 34 & 227 & 227/227 & 34 & 7 files / 43 signatures \\
DINOv3-ViT-H+/16 & 170 & 49 & 384 & 384/384 & 49 & 7 files / 51 signatures \\
\bottomrule
\end{tabular}
\end{table*}

The planner constructs candidates using fusion, schedule, memory, and parallel sub-planners. Each candidate specifies an implementation family and realizable parameters: fusion boundary, template type, tile size, thread count, vector width, reduction mapping, shared-memory policy, and launcher interface. Candidate generation is deliberately bounded: every plan must have a known template, a compatible launcher signature, and hardware-valid parameters before code generation. The code generator includes templates for elementwise chains, reductions, softmax, normalization, GEMM, Tensor-Core GEMM, GEMV, and selected fused epilogues. Validation combines compilation, interface checks, empirical numerical checks when available, and optional memory-safety tooling. The benchmark selector records latency for validated candidates and the emitter writes source, headers, wrappers, metadata, and build scripts.

\section{Kernel-Level Quality of LLM-Generated Kernels}
\label{app:kernel-results}

Table~\ref{tab:kernel-level} reports the best validated kernel per class, measured in-graph against the existing implementation it replaces. Two regimes emerge. Where fusion replaces a multi-op framework path, generated kernels win by large margins (3.7--4.3$\times$ for RoPE). Where the existing implementation is already near the bandwidth roofline, they cluster at parity to modest gains (0.96--1.12$\times$); the paged decode-attention kernel, which is generated from distilled principles alone and re-validated per model from a cross-model seed, reaches 0.955--0.975$\times$ of FlashAttention on four families (GLM-4's 32:2 grouped-query configuration reaches 0.45$\times$ and is rejected). Kernels that fail any gate or lose the measured comparison simply retain their baselines.

\begin{table}[t]
\centering
\small
\setlength{\tabcolsep}{4pt}
\caption{Best validated LLM-generated kernel per class, relative to the existing implementation it replaces, under the in-graph measurement protocol ($>$1 means the generated kernel is faster). The replaced implementations fall into three groups: (a) \emph{expert kernels}: attention preludes are measured against our hand-written fused reference kernels, and paged decode attention against FlashAttention; (b) \emph{library paths}, i.e., what production runs without the LLM tier: MLP rows against cuBLAS, and lm\_head+argmax against the cuBLAS + \texttt{torch.argmax} chain; (c) \emph{framework operator paths}: the RoPE row is measured against the multi-op framework path it replaces, a weaker baseline than group (a)/(b), which explains the large ratio. All MLP and lm\_head rows are the single-token ($M{=}1$) decode variants.}
\label{tab:kernel-level}
\begin{tabular}{llc}
\toprule
Kernel class & Best model & Ratio \\
\midrule
Attention prelude (QKV+RoPE) & Falcon3-7B & 1.03$\times$ \\
Attention prelude (+Q/K norm) & Qwen3-VL-2B & 1.02$\times$ \\
Paged decode attention & Qwen3-VL-2B & 0.975$\times$ \\
lm\_head+argmax ($M{=}1$) & Llama-3.2-3B & 1.07$\times$ \\
MLP gate--up + SiLU ($M{=}1$) & Qwen3-4B & 1.12$\times$ \\
MLP down + residual ($M{=}1$) & GLM-4-9B & 1.01$\times$ \\
Standalone RoPE & Qwen3-4B & 4.27$\times$ \\
\bottomrule
\end{tabular}
\end{table}

\section{Encoder-Only Vision Results}
\label{app:cv-results}

The pure-vision encoder DINOv3-ViT-H+/16 has no autoregressive decode, no KV cache, and no vLLM support, so we compare the assembled encoder, built from the same generation machinery through the encoder kernel family, against PyTorch eager across batch sizes (Table~\ref{tab:CV}). Gains are largest at small batches, where framework dispatch and small-kernel launch overhead dominate, and settle near parity as the workload becomes GEMM-bound; assembled features match the eager reference with cosine similarity 0.9998.

\begin{table*}[t]
\centering
\small
\setlength{\tabcolsep}{3.5pt}
\caption{End-to-end inference speedup of the assembled DINOv3 encoder over PyTorch eager. $B$ denotes the batch size; the last column reports the average over the eight batch sizes.} 
\label{tab:CV}
\begin{tabular}{lccccccccc}
\toprule
\multirow{2}{*}{Model} &
\multicolumn{8}{c}{Batch size $B$} &
\multirow{2}{*}{Avg. Speedup} \\
\cmidrule(lr){2-9}
& 1 & 2 & 4 & 8 & 16 & 32 & 64 & 96\\
\midrule
{\begin{tabular}[c]{@{}l@{}}dinov3-vith16plus\\-pretrain-lvd1689m\end{tabular}} & 2.57$\times$ & 1.93$\times$ & 1.11$\times$ & 1.06$\times$ & 1.03$\times$ & 1.03$\times$ & 1.02$\times$ & 1.02$\times$ & \textbf{1.35$\times$} \\
\bottomrule
\end{tabular}
\end{table*}

\section{Compiler-Pipeline Statistics}
\label{app:pipeline-evidence}
\paragraph{Analysis of Intermediate States}
Table~\ref{tab:pipeline-evidence} reports the compiler runs for all seven families. Each run traces the model into an internal IR, partitions it into regions, generates candidate CUDA implementations, validates them, benchmarks validated candidates, and emits deployment artifacts; executable kernels are thus selected through materialization, compilation, validation, benchmarking, and emission rather than textual plausibility. Every generated candidate passes validation (e.g., the Qwen3-4B log records maximum absolute and relative differences of 0.00e+00 across its 174 validations), one implementation is selected per region (selected-kernel latencies of 7.78--13.97~$\mu$s on Llama), and the DINOv3 assembled encoder matches the eager reference with feature cosine similarity 0.9998.

\section{Per-Model Adaptations}
\label{app:adaptations}

Table~\ref{tab:adaptations} summarizes the model-specific adaptations required to extend \textsc{AgentCompile} from its original Qwen3 and Llama models to seven model families. Most adaptations are one-time frontend or generator generalizations that benefit all subsequent models; structural gates ensure that unsupported patterns are skipped safely rather than mis-generated.

\begin{table*}[t]
\centering
\small
\setlength{\tabcolsep}{4pt}
\caption{Model-specific adaptations and their measured effects. Ratios are generated-kernel speed vs.\ the baseline implementation unless noted.}
\label{tab:adaptations}
\begin{tabular}{p{3.2cm}p{8.2cm}p{4.6cm}}
\toprule
Model & Adaptations & Measured effect \\
\midrule
Mistral-7B-Instruct-v0.3 & Sliding-window attention wired through the shared Llama-family container & Faster than vLLM on all measured workloads (1.08--1.11$\times$) \\
Falcon3-7B-Instruct & Architecture resolved from configuration metadata (fixing a silent wrong-class capture); Llama-family container reuse; per-model activation-budget calibration & Attention prelude 1.03$\times$; GEMV 1.04$\times$; zero OOM-driven concurrency throttling \\
GLM-4-9B-chat & Merged \texttt{gate\_up\_proj} layout adaptation (row-sliced zero-copy weight views feeding the fused MLP kernels); token-id clamping under configuration shrinking; dual defense for partial rotary (runtime width check + generator gate) & 53/53 kernels validated; MLP down+residual 1.01$\times$; end-to-end outputs match reference \\
Qwen3-VL-2B-Instruct & Text-only tracing for conditional-generation classes; recursive sub-configuration shrinking; text stack reuses the Qwen3 container & Q/K-norm RoPE prelude 1.02$\times$; GEMV 1.01$\times$ \\
DINOv3-ViT-H+/16 & Pure-vision encoder assembly layer: per-region numerical gating (cosine $\ge 0.9995$), bias folding, CUDA Graph replay with per-operator fallback & 51/51 kernels validated; assembled features cosine 0.9998 vs.\ eager \\
\bottomrule
\end{tabular}
\end{table*}

\section{Systems Compared}
\label{app:systems-compared}

\begin{table*}[t]
\centering
\small
\setlength{\tabcolsep}{3.1pt}
\caption{Runtime execution paths of compared systems during autoregressive generation (FA = FlashAttention).}
\label{tab:runtime-paths}
\begin{tabular}{lllll}
\toprule
System & Linear / MLP path & Attention path & KV cache & Decode control \\
\midrule
PyTorch Eager (FA) & cuBLAS & SDPA (flash backend) & DynamicCache & Python loop (eager) \\
\texttt{torch.compile}+FA & Inductor / Triton & FlashAttention & DynamicCache & Inductor graph \\
vLLM & cuBLAS + fused QKV & PagedAttention & Paged blocks & CUDA Graph replay \\
Ours (Single-Request)
& \makecell[l]{\texttt{CUDALinear} + generated GEMV +\\LLM-generated fused kernels} & \makecell[l]{LLM-generated paged decode\\attention (FA fallback)} & \makecell[l]{\texttt{FlashKVCacheManager}} & CUDA Graph replay \\
Ours (Multi-Request)
& Same as single-request & Same as single-request & \makecell[l]{Paged blocks +\\swap preemption} & \makecell[l]{Bucketed full-step\\CUDA Graph replay} \\
\bottomrule
\end{tabular}
\end{table*}

Table~\ref{tab:runtime-paths} summarizes the runtime path of each system. PyTorch eager uses HuggingFace/PyTorch generation with DynamicCache, framework dispatch, cuBLAS linear layers, and PyTorch SDPA when applicable. \texttt{torch.compile} with FlashAttention evaluates a PyTorch-native optimization path with Inductor fragments and a FlashAttention patch, but it still follows the framework generation path with dynamic KV cache growth. vLLM represents a serving runtime with PagedAttention, paged KV blocks, scheduling, and CUDA Graph replay for common decode shapes.

\textsc{AgentCompile} uses its own serving runtime: one-time preparation allocates the paged KV pool from a per-model memory budget, loads generated and baseline kernels under the model's name prefix, performs in-graph measured admission per kernel site, and captures bucketed full-step decode graphs. Timed prefill uses cuBLAS and FlashAttention (chunked when needed); timed decode replays the captured graph, executing admitted LLM-generated kernels with hand-written or library fallbacks at non-admitted sites; a use audit records the implementation serving each step.

\paragraph{Measurement protocol.}
The software stack includes HuggingFace Transformers, FlashAttention-2, Triton, vLLM, and CUDA compilation for generated kernels. Reported latency includes timed prefill and decode and excludes one-time preparation such as cache allocation, attention patching, extension loading, warmup, and CUDA Graph capture. Single-request timing uses two warmup iterations and ten measured repetitions. Workloads exceeding a model's maximum context length are excluded for all systems, since positional extrapolation makes such outputs meaningless.

\section{Multi-Request Workload Construction}
\label{app:workload-construction}

Each tier of the ragged swap suite (Table 4 in the main paper) is defined by a paged-KV pool size and a workload mix: the number in the tier label is the pool size in GB (e.g., 12M runs against a 12GB pool), and the letter denotes the mix. The base mix is a pool of mixed-length requests whose prompts range from a few hundred to about 4k tokens and whose outputs reach about 2k tokens. H (long) tiers scale all lengths by $3\times$, with prompts up to ${\sim}12$k tokens, and HH (extreme) tiers scale them by $5$--$6\times$; in both, the declared and actual lengths are equal. M (medium-pressure) tiers instead model early-stopping traffic: each request declares a large output budget but terminates early at its true generation length, and the pool is filled until the declared footprint (prompt plus declared maximum output) reaches about $3\times$ the KV-pool capacity, so the scheduler must over-commit, preempt, and swap based on declared rather than actual lengths. The maximum number of concurrently admitted requests is derived from the remaining GPU memory after subtracting the model weights, the KV pool, a safety margin, and a per-request activation estimate. Both systems serve the same request pool, that is, the identical prompts, length pairs, and arrival order, and we report end-to-end throughput over the whole pool. The uniform suite (Table 3 in the main paper) is the simpler counterpart: all requests in a tier share one input/output length, and the tier fixes the total number of requests and the maximum concurrency.

\section{Extended Related Work}
\label{app:related}

\textbf{Neural-network compilers:} TVM, XLA, PyTorch 2, MLIR, and Triton expose graph- or tensor-level abstractions for lowering models to accelerators~\citep{chen2018tvm,openxla2024xla,ansel2024pytorch,lattner2021mlir,tillet2019triton}; AutoTVM, Ansor, ROLLER, Welder, TASO, DNNFusion, AStitch, and Rammer automate schedules, rewrites, memory planning, or graph execution~\citep{chen2018learning,zheng2020ansor,zhu2022roller,shi2023welder,jia2019taso,niu2021dnnfusion,zheng2022astitch,ma2020rammer}. \textsc{AgentCompile} is complementary, adding an LLM-authored kernel tier on top of compiler-defined constraints, validation, and measurement.

\textbf{Inference runtimes and kernels:} On the serving side, vLLM and ORCA improve efficiency through scheduling, batching, and memory management~\citep{kwon2023efficient,yu2022orca}, with CUDA Graphs and FlashAttention providing launch-overhead and attention-kernel optimizations~\citep{nvidia2024cudagraphs,dao2022flashattention}; \textsc{AgentCompile} implements the same serving-grade mechanisms but sources its decode kernels from contract-gated LLM generation rather than a fixed expert-written library.

\textbf{LLMs for code and compiler optimization:} Among applications of LLMs to code and compiler optimization~\citep{chen2021evaluating,li2022competition,li2023starcoder,shinn2023reflexion,ni2024next,cummins2023large}, KernelBench evaluates direct LLM generation of GPU kernels~\citep{ouyang2025kernelbench}; \textsc{AgentCompile} differs in prompting from distilled principle libraries (including falsified dead ends) rather than task descriptions or reference source, and in requiring every kernel to pass compiler contracts, acceptance thresholds, and runtime measured admission with a never-worse fallback.

\bibliography{references}

\end{document}